\documentclass[journal,twocolumn]{IEEEtran}
\usepackage{epsfig,makeidx,color,epstopdf}
\usepackage{amsmath,amssymb,amsthm,bbm}
\usepackage{cite,graphicx}
\usepackage{enumerate}

\def\cK{{\cal K}}

\def\rH{{\rm H}}
\def\rT{{\rm T}}

\def\uE{{\mathbb E}}

\newtheorem{mylemma}{\bf Lemma} 

\def\be{ \begin{equation} }
\def\ee{ \end{equation} }
\def\bea{ \begin{eqnarray} }
\def\eea{ \end{eqnarray} }

\def\ba{{\bf a}}

\def\bm{{\bf m}}

\def\bR{{\bf R}}

\def\b0{{\bf 0}}

\def\cA{{\cal A}}

\def\cC{{\cal C}}

\def\cI{{\cal I}}

\def\cN{{\cal N}}

\ifCLASSOPTIONonecolumn
  \newcommand{\figwidth}{0.60\columnwidth}
\else
  \newcommand{\figwidth}{0.90\columnwidth}
\fi

\begin{document}

\title{NOMA based Random Access with Multichannel ALOHA}

\author{Jinho Choi\thanks{The author is with
School of Electrical Engineering and Computer Science,
Gwangju Institute of Science and Technology (GIST),
Gwangju, 61005, Korea (Email: \emph{jchoi0114@gist.ac.kr}).
This work was supported by
the ``Climate Technology Development and Application"
research project (K07732) through a grant provided by GIST in 2017.}}

\date{today}
\maketitle

\begin{abstract}
In nonorthogonal multiple access (NOMA),
the power difference of multiple signals is exploited
for multiple access and successive interference cancellation (SIC)
is employed at a receiver to mitigate co-channel interference.
Thus, NOMA is usually employed for coordinated transmissions
and mostly applied to downlink transmissions where 
a base station (BS) performs coordination for downlink
transmissions with full channel
state information (CSI).
In this paper, however, we show that
NOMA can also be employed for non-coordinated transmissions such as 
random access for uplink transmissions.
We apply a NOMA scheme to multichannel ALOHA and show that
the throughput can be improved. In particular,
the resulting scheme is suitable for random access 
when the number of subchannels is limited since
NOMA can effectively increase the number of subchannels
without any bandwidth expansion.
\end{abstract}

{\IEEEkeywords
random access; non-orthogonal multiple access;
throughput analysis}

\ifCLASSOPTIONonecolumn
\baselineskip 26pt
\fi

\section{Introduction}

Recently, nonorthogonal multiple access (NOMA)
has been extensively studied to improve the
spectral efficiency for future cellular systems, e.g., 
5th generation (5G) systems,
in \cite{Saito13, Kim13, Choi14, Ding15}.
In NOMA, a radio resource block is shared by multiple users
and their transmission power difference plays
a key role in multiple access.
Successive interference cancellation (SIC)
is also important in NOMA as it can mitigate co-channel interference
in a systematic manner.
In \cite{3GPP_MUST},
practical NOMA schemes, called
multiuser superposition transmission
(MUST) schemes, are considered for downlink transmissions
(with two users). In \cite{Choi14},
NOMA is employed for coordinated multipoint (CoMP) downlink
in order to support a cell-edge user without
degrading the spectral efficiency. 
For multiresolution broadcast, NOMA is studied with 
beamforming in \cite{Choi15}.
In \cite{Ding16A}, NOMA is also considered
for small packet transmissions
in the Internet of Things (IoT).

Machine-type communications (MTC) or
machine-to-machine (M2M) communications will play a crucial role
in 5G or the IoT \cite{Shar15} \cite{Bockelmann16}.
In \cite{Dhillon14} \cite{Wunder15}, it is shown that 
uncoordinated access or
random access schemes
might be suitable for MTC due to low signaling overhead
when devices have short packets to transmit. 
In general, random access for MTC is to provide access for uplink
transmissions (i.e., from devices to a base station 
(BS) or access point (AP)).

While NOMA has been actively studied 
as mentioned earlier, it is mainly considered for
downlink transmissions. Similarly, in the IoT
to support short packet transmissions, NOMA is employed
for downlink transmissions as in \cite{Ding16A}.
However, there are some existing works
of NOMA for uplink transmissions, e.g., \cite{Imari14}
\cite{Anxin15} \cite{Ding16_TWC}.

In general, NOMA requires coordinations with known
channel state information (CSI) to exploit the power difference
for multiple access.
Since the BS can carefully 
allocate powers to the signals to users with known CSI,
exploiting the power difference for NOMA becomes easier 
for downlink transmissions than uplink transmissions.
If NOMA is employed for uplink, the BS also needs to carefully
allocate powers and users over multiple channels
with full CSI as in  \cite{Imari14} \cite{Anxin15}.
From this, it seems that NOMA is not suitable for any
uncoordinated transmissions including random access despite
its strength of providing higher spectral efficiency.
In other words, NOMA may not be a suitable candidate for
random access to support MTC within 5G or the IoT.

In this paper, however, we consider NOMA for random access
where the BS does not perform any coordination for uplink transmissions.
In particular, we propose to apply a NOMA scheme to a well-known random
access scheme, multichannel ALOHA 
\cite{Yue91} \cite{Shen03}.
For the NOMA scheme,
we consider an approach in \cite{Choi16_DMA} that 
uses a set of pre-determined power levels for multiple access.
Using this NOMA
scheme, the throughput of multichannel
ALOHA can be improved without any bandwidth expansion.
The resulting scheme might be suitable for random access
when the number of subchannels is limited as 
NOMA can effectively increase the number of subchannels.
Consequently, when MTC is considered with a limited bandwidth,
the proposed scheme can be a good candidate for
random access due to more available subchannels.

Since the transmission power of the proposed random access scheme
based on NOMA can be high, we also study a channel-dependent selection
scheme for subchannel and power level, which can reduce
the transmission power.

The rest of the paper is organized as follows.
In Section~\ref{S:MA}, a well-known
random access scheme, multichannel ALOHA,
is briefly discussed.
In Section~\ref{S:PDMA}, a NOMA scheme
is presented as a random access scheme.
This NOMA scheme is applied to multichannel ALOHA
in Section~\ref{S:NMA} to effectively increase the number of subchannels
by exploiting the power domain.
Simulation results are presented in Section~\ref{S:Sim}.
The paper is concluded with some remarks in Section~\ref{S:Concl}.

\subsubsection*{Notation}
Matrices and vectors are denoted by upper- and lower-case
boldface letters, respectively.
The superscripts $*$, $\rT$, and $\rH$
denote the complex conjugate, transpose, Hermitian transpose, respectively.
For a set $\cA$, $|\cA|$ denotes the cardinality of $\cA$.
$\uE[\cdot]$ and ${\rm Var}(\cdot)$
denote the statistical expectation and variance, respectively.
$\cC \cN(\ba, \bR)$
represents the distribution of
circularly symmetric complex Gaussian (CSCG)
random vectors with mean vector $\ba$ and
covariance matrix $\bR$.

\section{Multichannel ALOHA}	\label{S:MA}

In this section, we 
briefly discuss multichannel (slotted) ALOHA
for uplink transmissions and its throughput.
Throughout the paper, we assume a single cell
with one BS and multiple users.

Multichannel ALOHA is a generalization of ALOHA with
multiple orthogonal subchannels \cite{Yue91} \cite{Shen03}.
In \cite{YJChoi06},
multichannel ALOHA is studied 
with orthogonal frequency division multiple access
(OFDMA) where each subcarrier becomes an orthogonal subchannel.

Suppose that there are $B$ orthogonal subchannels.
Denote by $\cI_i$ the index set 
of active users transmitting signals 
through the $i$th subchannel.
Then, the received signal at the BS
over the $i$th subchannel can be written as
\be
y_i = \sum_{k \in \cI_i} h_{i,k} \sqrt{P_{i,k} } s_{i,k} + n_i,
\ee
where $h_{i,k}$, $P_{i,k}$,
and $s_{i,k}$ represent the channel coefficient,
transmit power, and signal from user $k$ through
the $i$th subchannel, respectively, 
and $n_i \sim \cC \cN(0, N_0)$ is the background noise.
Here, $N_0$ is the noise spectral density.

Although it may be possible for the BS to detect some users'
signals when multiple users choose the same subchannel
due to the capture effect \cite{Yue91},
we ignore this possibility and employ a simple collision 
model \cite{BertsekasBook} for throughput analysis.
In this case, if there are $M$ active users
and each active user chooses a subchannel independently
and uniformly at random,
the conditional throughput\footnote{The throughput
is the average number of users who can successfully access
a channel without collision.} 
can be written as
\be
\eta_{\rm MA} (M;B) = M \left(1 - \frac{1}{B} \right)^{M-1}.
	\label{EQ:e1}
\ee

In \eqref{EQ:e1}, since $M$ is a random variable,
in order to find the average throughput,
we need to consider a distribution of $M$.
For convenience, we consider a uniform distribution with a large
number of users in this paper. To this end,
assume that there are $K$ users and each user
becomes active with access probability $p_{\rm a}$.
In addition, let $N$ denote the number of active users
that choose a subchannel.
Then, $\uE[M] =  K p_a$ and $\uE[N] = \frac{K p_a}{B}$.
For a large $K$,
we can use the Poisson approximation \cite{Mitz05}
for $N$.
That is,
$N$ becomes a Poisson random variable as follows:
\be
N \sim p_\lambda (n)  =
\frac{e^{-\lambda} \lambda^n}{n!},
\ee
where $p_\lambda (n)$ denotes the probability
mass function (pmf) of
a Poisson random variable with parameter $\lambda$. Here, 
$\lambda = \frac{\uE[M]}{B} = \frac{K p_{\rm a}}{B}$
is assumed to be constant,
which is called the intensity, as $K \to \infty$.
Then, the average throughput of multichannel ALOHA
can be found as
\begin{align}
T_{\rm MA} (B) & = \uE[\eta_{\rm MA} (M;B) ] \cr
& = B \lambda e^{-\lambda},
\end{align}
which is $B$ times higher than that of single-channel ALOHA
(with $B = 1$).
The intensity that maximizes
the throughput is $\lambda = 1$ 
\cite{BertsekasBook}
and the maximum
throughput is $B e^{-1}$.

\section{Random Access based on NOMA}	\label{S:PDMA}

In this section, we only consider a single subchannel
to present a random access scheme based on a NOMA scheme
studied in \cite{Choi16_DMA} and derive its 
conditional throughput.
For simplicity, we omit the subchannel index $i$
throughout this section.

\subsection{A NOMA Scheme: Power Division Multiple Access}

In this subsection, we consider a NOMA scheme that is suitable
for random access,
which is different from conventional uplink NOMA that
requires central
coordination including power allocation
at the BS with full CSI such as the approach in \cite{Imari14}.

Throughout the paper, we assume
that each user knows its CSI.
In time division duplexing
(TDD) mode, the BS can send a beacon signal at the beginning
of a time slot to synchronize uplink transmissions.
This beacon signal can be used
as a pilot signal to allow each user to estimate the CSI.
Due to various channel impairment (e.g., fading)
and the background noise,
the estimation of CSI may not be perfect.
However, for simplicity, we assume that the CSI estimation
is perfect in this paper. The impact of CSI estimation error 
on the performance needs to be studied in the future.
Suppose that there are pre-determined $L$ power levels
that are denoted by 
\be
v_1 > \ \ldots \ > v_L > 0.
	\label{EQ:vv}
\ee
We now assume that
an active user, say user $k$, can randomly choose
one of the power levels, say $v_l$, for 
random access. Then, the transmission power
is decided as
\be 
P_k = \frac{v_l}{\alpha_k},
	\label{EQ:CI}
\ee
where $\alpha_k = |h_k|^2$ is the channel power gain from
user $k$ to the BS, so that the
received signal power becomes $v_l$.
Assuming that the spectral density
of the background noise is normalized, i.e., $N_0 = 1$,
if there are no other active users, the 
signal-to-noise ratio (SNR) or 
signal-to-interference-plus-noise ratio (SINR) at the BS becomes
$v_l$.

Suppose that each power level in
\eqref{EQ:vv} is decided as follows:
\be
v_l = \Gamma (V_l + 1),
	\label{EQ:vV}
\ee
where $\Gamma$ is the target SINR and $V_l = \sum_{m=l+1}^L v_m$
with $V_L = 0$.
The value of the target SINR, $\Gamma$, can be decided depending
on the desired transmission rate or quality of link. 
It can be shown that 
\be
v_l = \Gamma (\Gamma +1 )^{L-l}.
	\label{EQ:vl}
\ee
If there exists one active user at each power level, the
SINR for the active user who chooses $v_1$
becomes $\frac{v_1}{V_1 + 1}$, which is $\Gamma$
from \eqref{EQ:vV}.
Thus, when the transmission rate, denoted by $R$,
is given by 
\be
R = \log_2(1 + \Gamma), 
\ee
the signal from this user
can be decoded and removed using SIC.
The SINR for the active user who chooses $v_{2}$ is also
$\Gamma = \frac{v_{2}}{V_2 + 1}$.
Consequently, all the $L$ signals can be decoded using
SIC in ascending order
if the transmission rate is given by $R = \log_2 (1+\Gamma)$.
In other words, a total of $L$ signals can be decoded
although they are transmitted simultaneously.

We can also observe that if there are $M$ active users with
$M \le L$ and they choose different power levels,
the $M$ signals can be successfully decoded.
This approach is referred to
as power-domain multiple access (PDMA) \cite{Choi16_DMA},
which can be seen as a NOMA scheme as the power domain
is exploited for multiple access and SIC is used to mitigate the 
co-channel interference.

Note that the above approach is based on ideal SIC
with capacity achieving codes. In practice, there might be decoding
errors and SIC may not be perfectly carried out. Thus,
a large $L$ is not desirable due to the error propagation.

\subsection{Throughput Analysis}

With the random access scheme based on
PDMA, the BS can successfully decode all signals from $M$ active
users if $M \le L$ and different powers are chosen.
However, if there are
multiple active users who choose the same power level,
the signals cannot be decoded.
For convenience, the event that
multiple active users choose the same power level
is called power collision.
Unlike conventional multichannel random access
schemes, the power collision at each power level
is not an independent event. That is,
if power collision happens at level $l$, the signals
at levels $l+1, \ldots, L$ cannot be decoded,
while the signals in the signals at higher power levels
can be decoded if there is no power collision.
For example, suppose that $L = 4$ and $M = 3$.
If one user chooses $v_1$ and the other two users choose
$v_4$, the signal from the user choosing $v_1$ can be decoded,
although the signals from the other two users cannot be decoded.

For the performance of random access
based on PDMA, we consider the conditional throughput
that is the average number of signals that are successfully decoded
for given $M$.
A bound on the (conditional) throughput
can be found as follows.

\begin{mylemma}	\label{L:1}
The conditional throughput for given $M \ (M \le L)$ active
users, denoted by $\eta(M;L)$,
 is bounded as
\begin{align}
\eta(M;L) 
& \ge {\underline \eta} (M;L) \cr
& = M \prod_{m=1}^{M-1} \left(1 - \frac{m}{L} \right).
	\label{EQ:BB}
\end{align}
If $M \le 2$, the bound is exact.
\end{mylemma}
\begin{IEEEproof}
The throughput,
${\underline \eta}(M;L)$
in \eqref{EQ:BB} corresponds to the case
that all $M$ signals can be decoded. Since the probability
that all $M$ signals have different power levels
is $\prod_{m=1}^{M-1} \left(1 - \frac{m}{L} \right)$
\cite{Mitz05}, we can have \eqref{EQ:BB}.
As mentioned earlier, since it is also possible to decode some signals 
in the presence of power collision, \eqref{EQ:BB}
becomes a lower-bound.

In the case of $M = 2$, the BS can decode
two signals if two active users choose different power levels.
If they choose the same power level, no signal can be decoded.
Thus, the lower-bound in \eqref{EQ:BB} becomes exact.
\end{IEEEproof}
Note that $\eta(M; L) \ge 0$ for any value of $M$.
Thus,
the lower-bound in \eqref{EQ:BB} is valid for any value of $M$
as ${\underline \eta}(M; L) =0$ for $M > L$.

\begin{figure}[thb]
\begin{center}
\includegraphics[width=\figwidth]{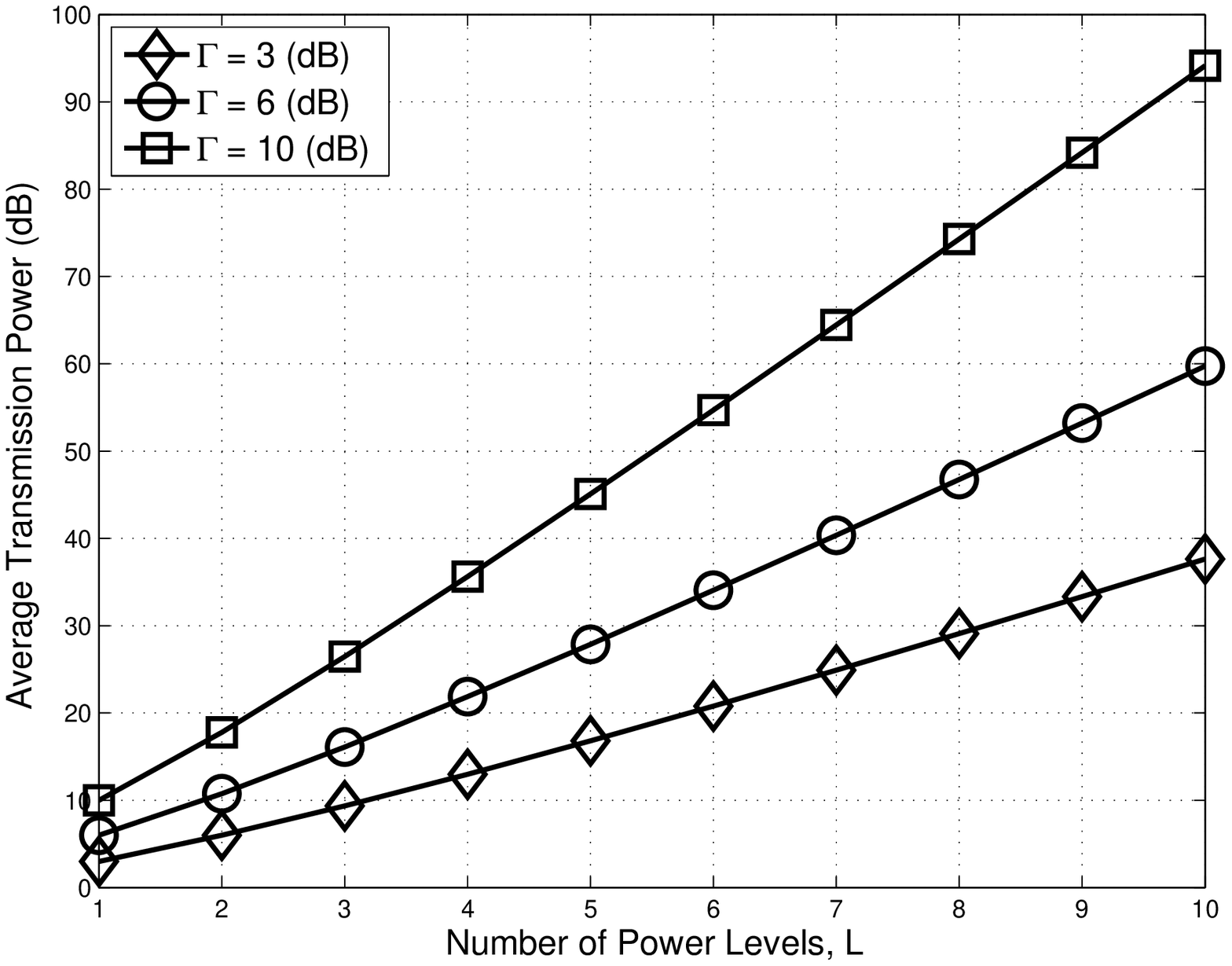}   \\
(a) \\
\includegraphics[width=\figwidth]{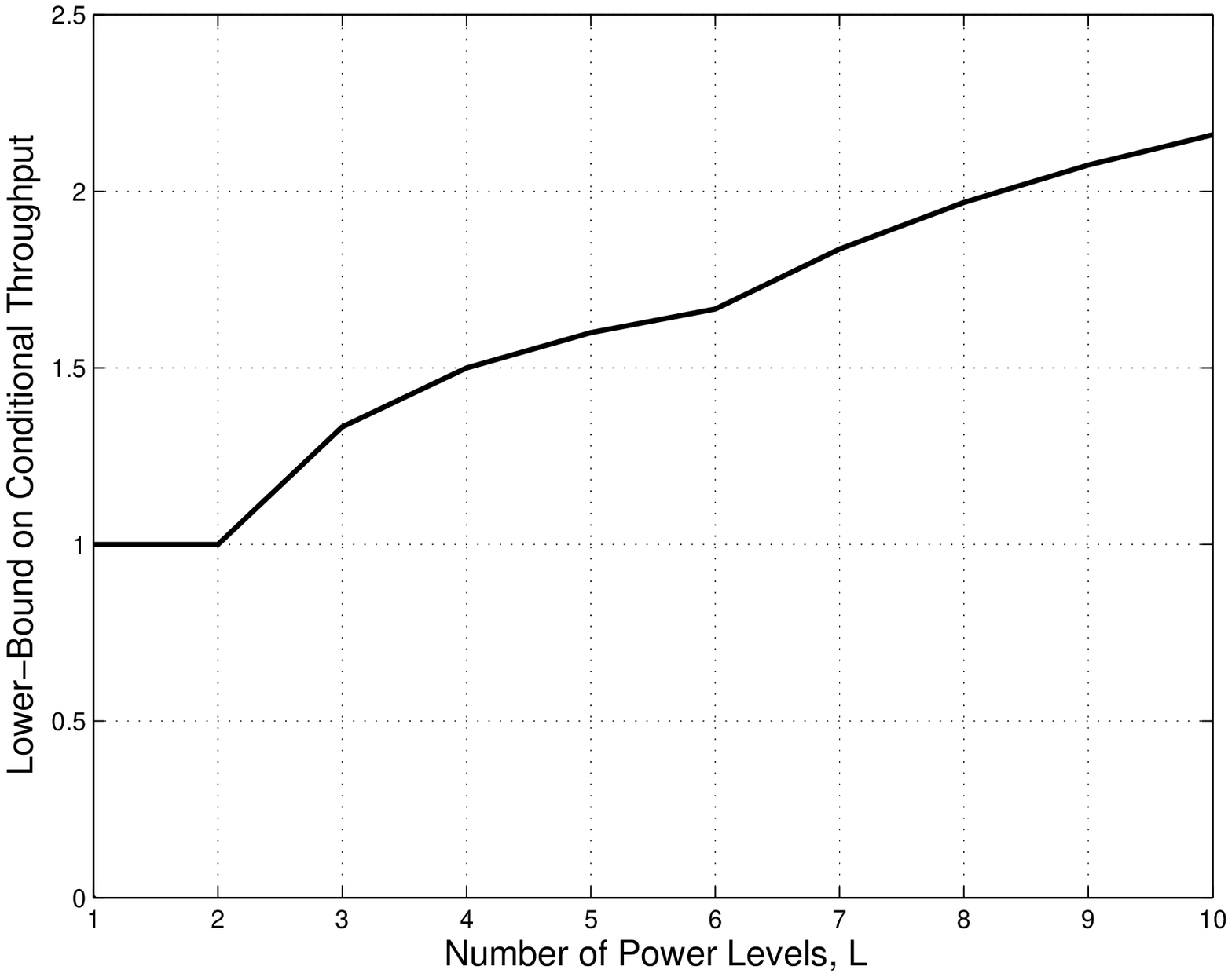}   \\
(b) 
\end{center}
\caption{Performance of PDMA for
different numbers of power levels, $L$:
(a) the average transmission power;
(b) the lower-bound on conditional throughput
(the optimal value of $M$ is chosen with
the lower-bound in \eqref{EQ:BB}).}
        \label{Fig:PL}
\end{figure}

From \eqref{EQ:BB}, we can show that
the resulting random access scheme based on PDMA 
can have a higher throughput
as $L$ increases. However, from
\eqref{EQ:vl}, since 
the highest power level,
$v_1 = \Gamma (\Gamma +1 )^{L-1}$,
grows exponentially with $L$,
a large $L$ becomes impractical.
In Fig.~\ref{Fig:PL}, we illustrate
the average transmission power 
of PDMA for different values of $L$
and target SINR. 
Fig.~\ref{Fig:PL} (a) 
shows the average transmission power of PDMA
for different numbers of power levels, $L$.
We can see that the increase of the
average transmission power is significantly higher
as $L$ increases. On the other hand,
the increase of the throughput with $L$ is not 
significant as shown in Fig.~\ref{Fig:PL} (b).
Thus, it may not be desirable to have a large $L$.
Note that the conditional throughput in 
Fig.~\ref{Fig:PL} (b) is the lower-bound in \eqref{EQ:BB}
where the optimal value of $M$ is chosen to maximize
the bound, i.e., $\max_{1 \le M \le L} {\underline \eta} (M;L)$.

\section{Application of NOMA to Multichannel ALOHA}	\label{S:NMA}

In this section, we 
propose a NOMA-multichannel ALOHA (NM-ALOHA) scheme by
applying PDMA to multichannel ALOHA, and study its 
throughput.
Furthermore, we study channel-dependent
selection for subchannel and power level to reduce the transmission power
or improve the energy efficiency.

\subsection{Application of PDMA to Multichannel ALOHA and
Throughput Analysis}

As discussed at the end of Section~\ref{S:PDMA},
random access based on PDMA may not be practical in terms of
its energy efficiency for a large $L$.
However, PDMA can be used with  other random access
schemes to improve throughput with a small $L$. In this subsection,
we consider NM-ALOHA using PDMA.

We assume that each subchannel in multichannel ALOHA
employs PDMA. Thus, there are $L B$ subchannels.
Suppose that there are $M_i$ active users 
that choose the $i$th subchannel.
In addition, let
$\bm = [M_1 \ \ldots \ M_B]^\rT$ 
and
$M = \sum_{i=1}^B M_i$.
Denote by $\eta_{\rm NMA} (\bm;L,B)$
the conditional throughput of NM-ALOHA using PDMA 
for given $\bm$.
Then, we have
\be
\eta_{\rm NMA} (\bm;L,B) =  \sum_{i=1}^B \eta(M_i;L).
\ee
\begin{mylemma}
If each active user can choose a subchannel uniformly at
random,
the conditional throughput for given $M$ 
can be found as
\begin{align}
T_{\rm NMA} (M;L,B) 
& = \uE[\eta_{\rm NMA} (\bm;L,B) \,|\, M] \cr
& = B \uE[\eta(N, L)],
	\label{EQ:b_eta}
\end{align}
where $N$ becomes the binomial random variable
with parameter $M$ and $p = \frac{1}{B}$, i.e.,
its pmf is given by
\begin{align*}
\Pr(N=n) & = p(n;M) \cr
& = \binom{M}{n} 
\left(\frac{1}{B} \right)^{n}
\left(1-\frac{1}{B} \right)^{M-n}.
\end{align*}
\end{mylemma}
\begin{IEEEproof}
It can be shown that
\begin{align}
T_{\rm NMA} (M;L,B) 
& = \sum_\bm \eta_{NMA} (\bm;L,B) p(\bm) \cr
& = \sum_\bm \sum_{i=1}^B \eta (M_i;L) p(\bm) 
\end{align}
where $p(\bm)$ is the pmf of multinomial random variables that
is given by
$p(\bm) = \frac{M}{M_1! \cdots M_B!} \left( \frac{1}{B} \right)^M$.
By marginalization, we can show that
$$
\sum_\bm \sum_{i=1}^B \eta (M_i;L) p(\bm) 
= B \sum_{n=1}^M \eta(n;L) p(n;M).
$$
Thus, we can have \eqref{EQ:b_eta}.
\end{IEEEproof}

For a large $K$, using the Poisson approximation,
from \eqref{EQ:BB} and \eqref{EQ:b_eta},
a lower-bound on the average throughput 
can be found as
\begin{align}
T_{\rm NMA} (L,B) & = \uE[ T_{\rm NMA} (M;L,B) ] \cr
& \ge B \sum_{n=1}^L {\underline \eta}(n;L) p_\lambda(n) \cr
& = B \sum_{n=1}^L n \left(
\prod_{m=1}^{n-1}  \left(1 - \frac{m}{L} \right)  \right)
\frac{e^{-\lambda} \lambda^n}{n!}.
	\label{EQ:LBT}
\end{align}
If $L = 2$,
as mentioned in Lemma~\ref{L:1},
the lower-bound is exact (because $M \le L$). 
Thus, 
we can show that
\begin{align*}
T_{\rm NMA} (2,B)
& = B \left( e^{-\lambda} \lambda + \frac{e^{-\lambda} \lambda}{2!}
\right) \cr
& = \frac{3}{2} B \lambda e^{-\lambda}.
\end{align*}
In addition, if $L = 1$, NM-ALOHA is reduced to 
standard multichannel 
ALOHA that has the following throughput:
$$
T_{\rm NMA} (1,B) = T_{\rm MA} (B) =  B \lambda e^{-\lambda}.
$$
From this, we can see that the average throughput
of NM-ALOHA with $L = 2$ is $1.5$ times higher than
that of standard multichannel ALOHA.
Furthermore, as 
${\underline \eta}(n;L)$ increases with $L$,
the lower-bound on
the average throughput increases with $L$.
Consequently, we can see that 
NM-ALOHA can improve the throughput
of multi-channel ALOHA 
without any bandwidth expansion
based on the notion of NOMA.

However, the increase of $L$
results in the increase of transmission power.
To mitigate the increase of transmission power,
we can consider a channel-dependent subchannel/power-level 
selection scheme
in the following subsection.

\subsection{Channel-Dependent Energy Efficient Selection}	\label{SS:CD}

In above, we assume that each active
user chooses a subchannel and a power level
independently and uniformly at random.
The selection of subchannel and power level
can depend on the channel gain and it may result in
the improvement in terms of energy efficiency
(or the decrease of transmission power).

Suppose that 
users are uniformly distributed within a cell of radius
$D$. 
We assume that
the large-scale fading coefficient of user $k$ is given by
\cite{TseBook05}
\be
\uE[\alpha_{i,k}] = \bar \alpha_k = A_0 d_k^{-\kappa},
\ 0 < d_k \le D,
	\label{EQ:LF}
\ee
where $\alpha_{i,k} = |h_{i,k}|^2$, $\kappa$ is the path loss exponent,
$A_0$ is constant, and $d_k$ is the distance between
the BS and user $k$. Thus, 
the large-scale fading coefficient depends on the distance.

For illustration purposes,
suppose that $L = 2$.
According to the large-scale fading coefficients or distances, 
we can divide users into two groups as follows:
\begin{align*}
\cK_1 & = \{k\,|\, d_k \le \tau \} \cr
\cK_2 & = \{k\,|\, d_k > \tau \}.
\end{align*}
If an active user belongs to $\cK_1$, this user
selects $v_1$. Otherwise, the user selects $v_2$.
That is, a user located far away from the BS
tends to choose a smaller $v_l$ to reduce the overall
transmission power.
We may decide the threshold value $\tau$ to satisfy the following
condition:
$$
\uE[|\cK_1|] = \uE[|\cK_2|] = \frac{K}{2},
$$
so that each group has the same number of users on average.
In this case, we have $\tau = \frac{D}{\sqrt{2}}$.
Consequently, the large-scale fading coefficient
is used as a random number for the power level
selection and the value of $\tau$ is decided
to make sure that $\Pr(k \in \cK_l) = \frac{1}{2}$,
(i.e., for a uniform power level selection at random). 

The above approach can be generalized for 
$L \ge 2$. To this end,
let 
\be
\cK_l =  \{k\,|\, \tau_{l-1} < d_k \le \tau_l \}.
\ee
Under the assumption that users are uniformly distributed
in a cell of radius $D$,
we have
$\tau_0 = 0$ and $\tau_l = D \sqrt\frac{l}{L}$,
$l = 1,\ldots,L$, to satisfy
$$
\Pr(k \in \cK_l) = \frac{1}{L} , \ l = 1,\ldots, L,
$$
which also results in
$\uE[|\cK_l|] = \frac{K}{L}$. 
To minimize the transmission power,
an active user belongs to $\cK_l$ chooses $v_l$.

Furthermore, when an active user in $\cK_l$ chooses
one of $B$ subchannels in NM-ALOHA, the user may
choose the subchannel that has the maximum
channel gain to further minimize the transmission power.
As a result, the transmission
power of user $k$ can be decided as
\be
P_k = \frac{v_l}{\max_i \alpha_{i,k}}, \ k \in \cK_l.
	\label{EQ:P_kK}
\ee
Note that in this case, if $\alpha_{1,k}, \ldots, \alpha_{B,k}$
are independent and identically distributed 
(iid), the selection of subchannel 
is carried out independently and uniformly at random.
The selection scheme resulting in 
\eqref{EQ:P_kK} is referred to as
the channel-dependent subchannel/power-level selection scheme.

\begin{mylemma}
Suppose that 
\be
\alpha_{i,k} = \bar \alpha_k u_{i,k}^2,
	\label{EQ:SF}
\ee
where $u_{i,k}$ is an independent Rayleigh random variable
with $\uE[u_{i,k}^2] = 1$
(i.e., small-scale fading is assumed to be Rayleigh distributed).
Then, for $B \ge 2$,
the average transmission power is bounded as
\begin{align}
\uE[P_k\,|\, k \in \cK_l] 
\le \frac{v_l}{A_l} \min\left\{ 2 \ln 2,\frac{B}{B-1} \right\},
	\label{EQ:EP}
\end{align}
where $A_l = A_0 \tau_l^{-\kappa}$.
\end{mylemma}
\begin{IEEEproof}
To find an upper-bound, we consider
a user of the longest distance within $\cK_l$, $\tau_l$. 
In this case, we have
\be
\alpha_{i,k} = A_l u_{i,k}^2, \ i = 1, \ldots, B,
\ee
which are iid.
According to order statistics \cite{DavidBook},
we can see that 
$\uE\left[ \frac{1}{\max_i \alpha_{i,k}} \right]$
is a nonincreasing function of $B$.
Thus, for an upper-bound, it is sufficient to consider
the case of $B = 2$.
Since $u_{i,k}^2$ is an exponential random variable,
it can be shown that
\begin{align}
\uE\left[ \frac{1}{\max \alpha_{i,k}} \right]
& = \frac{1}{A_l} 
\int_0^\infty \frac{1}{x} B e^{-x} (1- e^{-x})^{B-1} dx \cr
& \le \frac{2}{A_l} 
\int_0^\infty \frac{1}{x} e^{-x} (1- e^{-x}) dx \cr
& = \frac{2 \ln 2}{A_l}, 
	\label{EQ:1max}
\end{align}
where the last step is due to \cite[Eq. (3.434)]{Gradshteyn}.

To find another bound for any $B \ge 2$, 
let $t = 1  - e^{-x}$. Then, it can be shown that
\begin{align}
\int_0^\infty \frac{1}{x} B e^{-x} (1- e^{-x})^{B-1} dx 
& = B \int_0^1 -\frac{t^{B-1}}{\ln(1-t)} dt \cr
& \le B \int_0^1 \frac{t^{B-1}}{t} dt \cr
& = \frac{ B}{B-1},
	\label{EQ:BB1}
\end{align}
where the inequality is due to $t \le - \ln (1-t)$, $t \in (0,1)$.
Substituting \eqref{EQ:BB1} into \eqref{EQ:1max},
we have
\begin{align}
\uE\left[ \frac{1}{\max \alpha_{i,k}} \right]
\le \frac{1}{A_l} \frac{B}{B-1}.
	\label{EQ:2max}
\end{align}
From \eqref{EQ:1max} and \eqref{EQ:2max}, 
we can readily show \eqref{EQ:EP}.
\end{IEEEproof}

From \eqref{EQ:vl} and \eqref{EQ:EP}, noting that
$\Pr(k \in \cK_l) =\frac{1}{L}$,
the average transmission power is upper-bounded as
\begin{align}
\uE[P_k] 
& \le \frac{\min\left\{2 \ln 2, \frac{B}{B-1} \right\}
}{L} \sum_{l=1}^L \frac{v_l}{A_l} \cr
& = \frac{\min\left\{2 \ln 2, \frac{B}{B-1} \right\}
}{L} \sum_{l=1}^L
\frac{\Gamma (\Gamma +1)^{L-l}}{A_0 
\left(D  \sqrt\frac{l}{L} \right)^{-\kappa}}.
	\label{EQ:UTP}
\end{align}

It is noteworthy that under \eqref{EQ:SF},
$\uE \left[\frac{1}{\alpha_{i,k}} \right] \to \infty$,
which is the case of $B = 1$.
Thus, the power allocation in \eqref{EQ:P_kK} 
may result in a prohibitively high transmission power.
To avoid this problem, the truncated channel inversion power
control can be used \cite{Goldsmith97a}.
However, if $B \ge 2$, this problem can be mitigated
without any transmission power truncation,
since $\uE \left[\frac{1}{\alpha_{i,k}} \right] < \infty$
from \eqref{EQ:1max}.

For comparison purposes,
we now consider a random selection for subchannel
and power level.
If the subchannel and power level are randomly selected,
the average transmission power 
would be
\begin{align}
\uE[P_k] 
& = \frac{1}{L} \sum_{l=1}^L \uE \left[\frac{v_l}{\alpha_{i,k}}
\right] \cr
& = \frac{1}{L} \sum_{l=1}^L v_l  
\uE \left[\frac{1}{\alpha_{i,k}} \right].
\end{align}
In this case, even if  
$\uE \left[\frac{1}{\alpha_{i,k}} \right]$ converges to constant,
we note that
\be
\uE[P_k] 
\propto \frac{1}{L} \sum_{l=1}^L v_l.
	\label{EQ:RPow}
\ee
Thus, from \eqref{EQ:RPow} and \eqref{EQ:UTP}, we can see that
the average transmission power with 
channel-dependent (subchannel/power-level) selection  
grows slower than that with 
random selection as $L$ increases.
This shows the advantage of 
channel-dependent selection  over random selection
for the selection of subchannel and power level
in NM-ALOHA in terms of the transmission power for
$B > 1$ and $L > 1$.

It is noteworthy that the channel-dependent selection
does not affect the throughput as users' locations are random,
while it can greatly improve the energy efficiency.


\section{Simulation Results}	\label{S:Sim}

In this section, we present simulation results
to see the performance of NM-ALOHA
with the fading channel coefficients, $h_{i,k}$, that are generated
according to \eqref{EQ:LF} 
and \eqref{EQ:SF}.
For the path loss exponent, $\kappa$, in 
\eqref{EQ:LF}, we assume that $\kappa = 3.5$.
In addition, we assume that $D = 1$ and $A_0 = 1$ in
\eqref{EQ:LF} for normalization
purposes.

\begin{figure}[thb]
\begin{center}
\includegraphics[width=\figwidth]{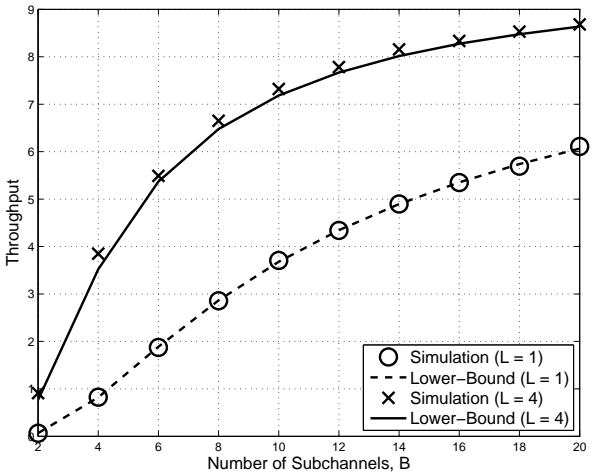}  
\end{center}
\caption{Throughput of NM-ALOHA
for different number of subchannels
when $K = 200$, $p_a = 0.05$, and
$L \in \{1, 4\}$.}
        \label{Fig:aplt3}
\end{figure}

Fig.~\ref{Fig:aplt3}
shows the throughput of NM-ALOHA
for different numbers of subchannels, $B$,
when $K = 200$, $p_a = 0.05$, and
$L \in \{1, 4\}$.
The lower-bound is obtained from \eqref{EQ:LBT}.
As expected, we can observe that
the throughput increases with the number of subchannels, $B$.
More importantly, we can see that 
the throughput of NM-ALOHA ($L = 4$) is higher than
that of (conventional) multichannel ALOHA ($L = 1$).
In particular, when $B = 4$,
the throughput of 
NM-ALOHA becomes about 4 times higher than that of
multichannel ALOHA,  while the throughput gap
decreases with $B$.
This demonstrates that when the number of subchannels
is limited in multichannel ALOHA, NOMA
based approaches such as
NM-ALOHA can help improve the throughput.
For example, NM-ALOHA ($L = 4$) can achieve
a throughput of 3.5 with $B = 4$, while the same
throughput can be obtained by multichannel ALOHA with 
$B = 10$ (in this case, we can claim that NM-ALOHA can be 2.5
times more spectrally efficient than multichannel ALOHA).

\begin{figure}[thb]
\begin{center}
\includegraphics[width=\figwidth]{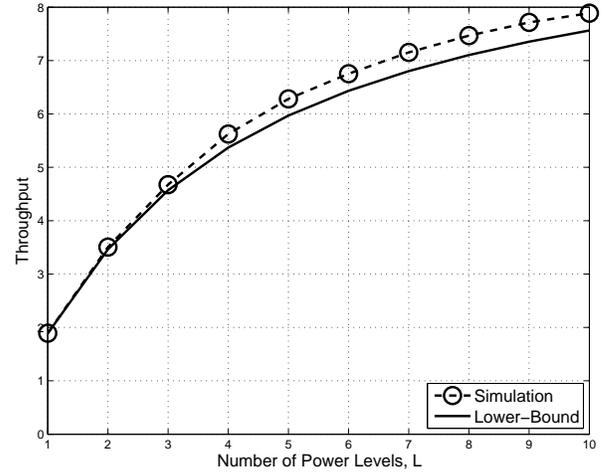}  
\end{center}
\caption{Throughput for 
different values of $L$ 
when $K = 200$, $p_a = 0.05$, and $B = 6$.}
        \label{Fig:aplt1}
\end{figure}

In order to see the impact of
the number of power levels, $L$, on the 
throughput of NM-ALOHA, we 
show the throughput for 
different values of $L$ in 
Fig.~\ref{Fig:aplt1} 
when $K = 200$, $p_a = 0.05$, and $B = 6$.
As expected, the throughput increases
with $L$ without any bandwidth expansion
(i.e., with a fixed $B$).
For example, with $L  = 4$, the throughput can be 3 times
higher than that of (conventional) multichannel ALOHA 
(i.e., NM-ALOHA with $L = 1$).
However,
the improvement of throughput becomes limited
when $L$ is sufficiently large.

\begin{figure}[thb]
\begin{center}
\includegraphics[width=\figwidth]{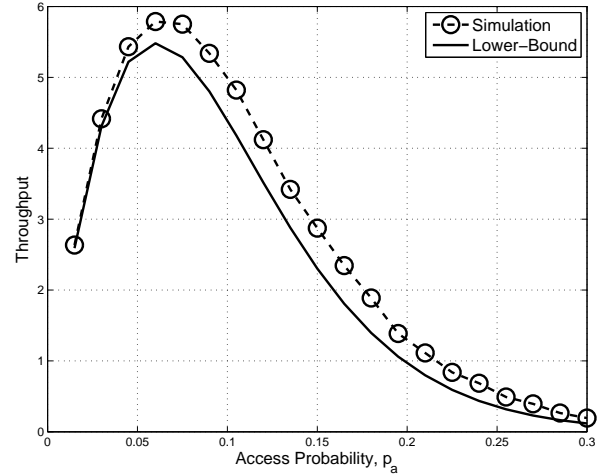}
\end{center}
\caption{Throughput for different values of
access probability, $p_a$,
when $K = 200$, $L = 4$, and $B = 6$.}
        \label{Fig:aplt2}
\end{figure}

In Fig.~\ref{Fig:aplt2},
we show the throughput for different values of
access probability, $p_a$,
when $K = 200$, $L = 4$, and $B = 6$.
The performance behavior of NM-ALOHA is similar
to that of multichannel ALOHA in terms of
$p_a$. That is, the throughput increases with $p_a$,
and then decreases,
which implies that there exists an optimal
access probability that maximizes the throughput.
Thus, it is possible to consider the access control
using the access probability
as in ALOHA \cite{BertsekasBook} or the number of
subchannels \cite{Choi_NC}.
However, this topic is beyond the scope of the paper and might
be further studied in future research.

From Figs.~\ref{Fig:aplt3} -- \ref{Fig:aplt2},
we can confirm that the lower-bound 
from \eqref{EQ:LBT}
is reasonably tight,
while it becomes tighter as $L$ decreases 
as shown in Fig.~\ref{Fig:aplt1}.

\begin{figure}[thb]
\begin{center}
\includegraphics[width=\figwidth]{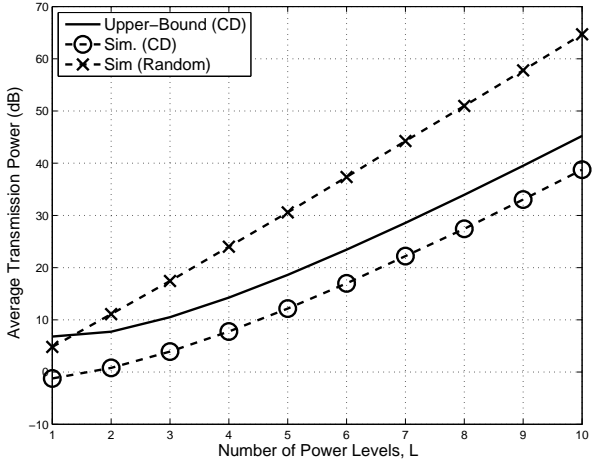}  
\end{center}
\caption{Average transmission power for 
different values of $L$
when $K = 200$, $p_a = 0.05$, $B = 6$,
and $\Gamma = 6$ dB 
(in the legend, `CD'
represents the result obtained by the channel-dependent selection).}
        \label{Fig:Tplt1}
\end{figure}

The main disadvantage of NM-ALOHA might be 
a high transmission power as mentioned earlier.
To mitigate this problem, we considered
the channel-dependent (subchannel/power-level) selection
scheme in Subsection~\ref{SS:CD}.
To see the impact of this selection 
scheme on the average transmission power,
we present simulation
results in Fig.~\ref{Fig:Tplt1}
where the average transmission power
is shown for different values of $L$
when $K = 200$, $p_a = 0.05$, $B = 6$,
and $\Gamma = 6$ dB. We also show
the upper-bound in \eqref{EQ:UTP}.
Furthermore, for performance comparisons,
we consider the random selection
for subchannel and power level regardless
of the channel conditions.
Since the transmission power can be arbitrarily high
due to the channel inversion power control
in \eqref{EQ:CI}, we assume that the transmission
power is limited to be less than or
equal to $10 L$ dB 
(i.e., truncated power control is assumed)
in simulations hereafter.
The corresponding results are shown with the legend 'Sim (Random)'
in Fig.~\ref{Fig:Tplt1}.
We can observe that the average transmission power
increases with $L$, while 
the channel-dependent selection
scheme provides a much lower 
average transmission power
than the (channel-independent) random selection scheme.

\begin{figure}[thb]
\begin{center}
\includegraphics[width=\figwidth]{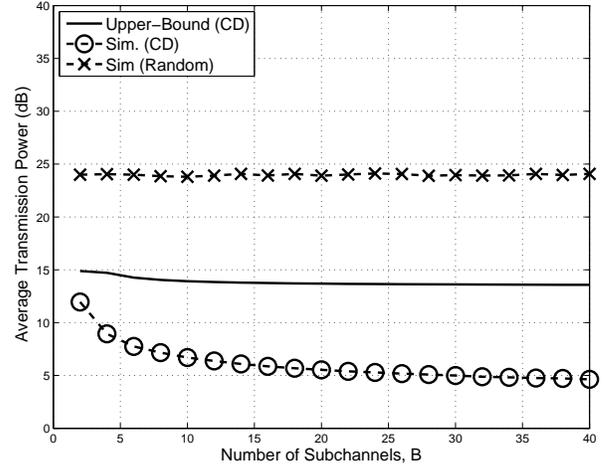}  
\end{center}
\caption{Average transmission power for 
different values of $B$
when $K = 200$, $p_a = 0.05$, $L = 4$,
and $\Gamma = 6$ dB
(in the legend, `CD'
represents the result obtained by the channel-dependent selection).}
        \label{Fig:Tplt2}
\end{figure}

Fig.~\ref{Fig:Tplt2} shows 
the average transmission power for 
different numbers of subchannels, $B$,
when $K = 200$, $p_a = 0.05$, $L = 4$,
and $\Gamma = 6$ dB.
As expected, 
the average transmission power decreases with $B$.
On the other hand,
the average transmission power with the random selection
does not depend on $B$.
Consequently, we can see that 
although a large $B$ 
does not help improve the throughput significantly 
(with a fixed $p_a$)
in NM-ALOHA as shown in Fig.~\ref{Fig:aplt3},
it can be effective for improving energy efficiency
with channel-dependent selection.
Note that the upper-bound derived in \eqref{EQ:UTP}
is tight when $B$ is small.

\begin{figure}[thb]
\begin{center}
\includegraphics[width=\figwidth]{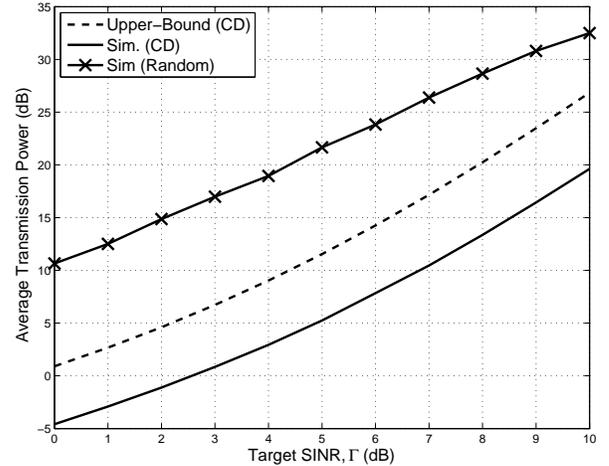}  
\end{center}
\caption{Average transmission power for 
different values of $\Gamma$
when $K = 200$, $p_a = 0.05$, $L = 4$, and $B = 6$.
(in the legend, `CD'
represents the result obtained by the channel-dependent selection).}
        \label{Fig:Tplt3}
\end{figure}

Fig.~\ref{Fig:Tplt3} shows 
the average transmission power for 
different values of the target SINR, $\Gamma$,
when $K = 200$, $p_a = 0.05$, $L = 4$, and $B = 6$.
We can see that the average transmission power increases with
$\Gamma$. Thus, the target SINR should be decided according
to a given feasible average transmission power.

\section{Concluding Remarks}	\label{S:Concl}

In this paper, we proposed a random access scheme 
by applying NOMA to multichannel ALOHA.
The proposed scheme has multiple subchannels
and multiple power levels for random access
to effectively increase the number of subchannels.
It was shown that
the proposed scheme can provide a higher
throughput than 
multichannel ALOHA by exploiting the power difference. 
As a result, the proposed scheme 
became suitable for random access 
when the number of subchannels of multichannel ALOHA is limited,
because
NOMA can effectively increase the number of subchannels
without any bandwidth expansion.
A closed-form expression for a lower-bound on the 
throughput was derived to see the performance.

The main drawback of the proposed scheme
was a high transmission power that is a typical problem of
NOMA as the power domain is exploited.
In order to mitigate this problem,
a channel-dependent 
selection scheme for subchannel and power level
was studied, which leads to the decrease of transmission power or
improvement of energy efficiency.
An upper-bound on the average transmission power was
derived to see the impact of the 
channel-dependent 
selection scheme on the average
transmission power in terms of the number of power levels.

\bibliographystyle{ieeetr}
\bibliography{noma}

\begin{IEEEbiography}[{\includegraphics[width=1in,height=1.25in,keepaspectratio]{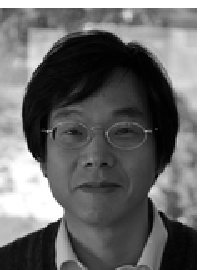}}]
{Jinho~Choi}
(SM'02) was born in Seoul, Korea. He received B.E. (magna cum laude) degree
in electronics engineering in 1989 from Sogang University, Seoul, and
M.S.E. and Ph.D. degrees in electrical engineering from
Korea Advanced Institute of Science and Technology (KAIST), Daejeon,
in 1991 and 1994, respectively.
He is with Gwangju Institute of Science and Technology (GIST) as a Professor.
Prior to joining GIST in 2013, he was with
the College of Engineering, Swansea University, United Kingdom,
as a Professor/Chair in Wireless.
His research interests include wireless communications and array/statistical
signal processing. He authored two books published by Cambridge University
Press in 2006 and 2010. Prof. Choi received
the 1999 Best Paper Award for Signal
Processing from EURASIP, 2009 Best Paper Award from WPMC (Conference),
and is Senior Member of IEEE. Currently, he is an Editor of
IEEE Trans. Communications and had served as an Associate Editor
or Editor of other journals
including IEEE Communications Letters, Journal of Communications
and Networks (JCN), IEEE
Transactions on Vehicular Technology, and ETRI journal.
\end{IEEEbiography}
\end{document}